# Electronic Duality in Strongly Correlated Matter

T. Park, M. J. Graf, L. Boulaevskii, J. L. Sarrao & J. D. Thompson

*Los Alamos National Laboratory, Los Alamos, NM 87545, USA*

**Superconductivity develops from an attractive interaction between itinerant electrons that creates electron pairs which condense into a macroscopic quantum state--the superconducting state. On the other hand, magnetic order in a metal arises from electrons localized close to the ionic core and whose interaction is mediated by itinerant electrons. The dichotomy between local moment magnetic order and superconductivity raises the question of whether these two states can coexist and involve the same electrons. Here we show that the single 4f-electron of cerium in $CeRhIn_5$ simultaneously produces magnetism, characteristic of localization, and superconductivity that requires itinerancy. The dual nature of the 4f-electron allows microscopic coexistence of antiferromagnetic order and superconductivity whose competition is tuned by small changes in pressure and magnetic field. Electronic duality contrasts with conventional interpretations of coexisting spin-density magnetism and superconductivity and offers a new avenue for understanding complex states in classes of materials.**

The possibility that the same electrons might exhibit simultaneously localized and itinerant characters has been raised in the context of materials in which strong Coulomb interactions nominate physical properties. $UPd_2Al_3$ is one such correlated electron material in which coexisting antiferromagnetism and superconductivity may be interpreted if two of uranium's three 5f-electrons are localized spatially close to the ionic core and produce antiferromagnetic order, whereas the remaining f-electron is spatially delocalized and participates in creating superconductivity (1). In elemental Pu metal, the volume and electronic spectrum of its δ-phase can be described *ad hoc* if one of Pu's five 5f-electrons were itinerant and four of the 5f-electrons were localized (2, 3). Theoretically, the competition between intra-atomic Coulomb interactions and anisotropic hybridization of f-electrons with their chemical environment is one potential route to the division of 5f-orbitals into localized and delocalized components (4). A more perplexing situation is presented if a single f-electron, such as in Ce, were to display localized and itinerant natures simultaneously. Like $UPd_2Al_3$, $CeRhIn_5$ is a strongly correlated metal in which antiferromagnetic order and superconductivity coexist (5), and as we will show, this coexistence necessitates the concept of electronic duality.

Within the resolution of electronic structure calculations and measurements, the single 4f-electron of Ce in $CeRhIn_5$ is localized (6), consistent with the observation of antiferromagnetic ordering of the nearly full magnetic moment carried by a localized $4f^1$ electron in a crystalline electric-field doublet ground state (7). A slight (~10%) reduction of the ordered moment from its full value is due to weak hybridization of the 4f electron with ligand electrons, which transfers some spectral weight of the f electron to itinerant band states. Applying a sufficiently high pressure (greater than a critical value P2 marked in Fig. 1a) to $CeRhIn_5$ promotes stronger hybridization such that the 4f electron becomes itinerant at low temperatures, contributes to the electronic band structure and participates in superconductivity (8, 9). Nuclear magnetic resonance and specific heat studies establish that the superconductivity is unconventional (10, 11): unlike conventional superconductors, the superconducting energy gap, which develops because itinerant electron form pairs, in $CeRhIn_5$ contains nodes where the gap becomes zero on parts of the Fermi surface. As also shown in Fig. 1a, antiferromagnetic order (AFM) and unconventional superconductivity (SC) coexist over a range of pressures below P1. Measurements of specific heat divided by temperature C/T, plotted in Fig. 1b, substantiate conclusions from nuclear magnetic resonance experiments that demonstrate homogeneous, microscopic coexistence of AFM and SC below P1 and the absence of magnetic order above P1 (ref. 10). The area under these C/T curves gives the electronic entropy. For temperatures less than ~13 K, the entropy is independent of pressure (11) and implies that the ground state, whether AFM, SC or a phase of coexisting AFM and SC, is controlled by the fate of the 4f electron that is revealed on an energy scale of 1 meV.

An expanded view of the specific heat divided by temperature C/T is plotted in Fig. 2a for a representative pressure where AFM and SC coexist. Application of a magnetic field at this pressure has little effect on the magnetic transition temperature $T_N$ but suppresses superconductivity to reveal a finite $T \rightarrow 0$ value of C/T, $\gamma_N$, that is more than three times larger than $\gamma_N$ at atmospheric pressure (5, 12). In the absence of an applied field, superconductivity emerges from electronic degrees of freedom measured by the magnitude of $\gamma_N$ but not all electronic states at the Fermi energy are gapped by superconductivity (13), as evidenced by a residual T=0 value of C/T, $\gamma(0)$. The evolution of $\gamma_N$ and $\gamma(0)$ with pressure is given in the inset of Fig. 2b, where we see that $\gamma(0)$ becomes negligibly small above P1, accompanying the collapse of magnetic order, but $\gamma_N$ increases as P approaches P1, exceeding 50% of C/T at temperatures just above the magnetic transition. Because at these pressures the ordered moment of the 4f electron is still at least 80% of its ambient-pressure value (14), this is strong indication that the 4f electron has assumed both localized and itinerant characters. Accompanying these changes is a pronounced increase near P1 in the normalized specific heat jump $\Delta C_{SC} / \gamma_N T_c$ at the superconducting transition temperature $T_c$ (see Fig. 2b) that implies a qualitative change in electronic structure with no change in crystal structure (15). Above P1, the large ratio $\Delta C_{SC} / \gamma_N T_c$ becomes comparable to that found in the isostructural and nominally isoelectronic unconventional superconductor $CeCoIn_5$ (ref. 16), whose itinerant 4f electron participates in superconductivity and creates a large Fermi volume similar to that of $CeRhIn_5$ at $P \geq P2$ (ref. 8). Below P1, the nearly constant value of this ratio implies that superconductivity is developing from the itinerant component of the 4f electron that is reflected in $\gamma_N$.

Measurements in a magnetic field explicitly reveal the dual nature of Ce's 4f electron and its role in creating coexisting phases. Figure 3 shows field-temperature phase diagrams for $CeRhIn_5$ at representative pressures where AFM and SC coexist. Except for the presence of superconductivity, these diagrams are very similar to one constructed at atmospheric pressure (17, 18). With increasing pressure, the field required to induce a change in magnetic structure increases even though $T_N(B=0)$ decreases. Competition between the field-induced commensurate (CM) magnetic phase and superconductivity reflects competing tendencies of the 4f electron: a higher field is required to suppress pressure-induced hybridization of f and ligand electrons that give the 4f electron its itinerant nature.

The upper critical field $B_{c2}$, above which superconducting state becomes normal, shows qualitatively different forms of scaling across P1 (Fig. 4a). For pressures above P1, where a solely SC phase reflects itinerancy of 4f electrons, polarization of the enhanced density of itinerant spins by an applied field strongly limits $B_{c2}$ as temperature approaches zero (19). In contrast, for pressures below P1, where coexisting AFM and SC reflects duality of 4f electrons, this limiting mechanism is absent and $B_{c2}$ increases approximately linearly with decreasing temperature. From the slope of $B_{c2}$ versus T near $T_c$, we can calculate the Fermi velocity ($v_F$) of itinerant electrons (20). As shown in Fig. 4b, $v_F$ decreases gradually as P approaches P1, signalling the increasing development of an itinerant component in the f electron, and a sudden drop at P1 marks a transition in electron structure. This transition is not observed in de Haas-van Alphen (dHvA) measurements because these experiments, which directly measure the itinerant electron band structure, are made at high fields which force a change from itinerant to localized behaviour. At these high fields, however, dHvA finds that the balance between localized and itinerant natures of the 4f electron is eventually tipped at P2 to itinerancy, which is favoured by high pressure (8). The two critical points P1 and P2, then, are Lifshitz-like transitions (21), one in zero-field (P1) and the other in high field (P2). Consequently, the nature of field-tuned quantum criticality observed (9) between P1 and P2 is a multi-critical line where the Fermi surface reconstructs and magnetic fluctuations diverge, a scenario anticipated by theory (22, 23).

The relationship between large-moment magnetism and superconductivity as a function of pressure and temperature shown in Fig. 3 as well as the small-to-large Fermi volume found at P2 (ref. 8) cannot be explained by conventional models that attribute both AFM and SC to an instability of a sea of itinerant electrons (24, 25). Electronic duality manifested in $CeRhIn_5$ requires a new conceptual framework that poses a challenge to theory. An appropriate description of strong electronic correlations must be a key ingredient of this framework. Though progress is being made (26), theoretical accessibility to essential low energy scales, such as found in $CeRhIn_5$, is missing but necessary to reveal duality and its consequences. A two-fluid phenomenology (27, 28), for which a theoretical basis is beginnng to emerge (29), appears to capture aspects of the low-energy duality in families of strongly correlated materials; however, this phenomenology does not include interactions that would lead to low temperature broken symmetries. On the other hand, complex gauge theories, such as SO(5), can account for coexisting magnetism and superconductivity, but its applicability is difficult to test experimentally (30). Entangled spin and charge degrees of freedom and associated complex states are not unique to $CeRhIn_5$ but are found as well in the high-transition temperature superconductors (31, 32). An appropriate description of electronic duality holds promise for a solution to both problems.

**Materials and Methods:**

Single crystals of CeRhIn$_5$, which were grown using In-flux technique (5), are of exceptionally high quality, having a residual electrical resistivity on the order of 40 n$\Omega$ cm. Hydrostatic pressure was achieved by using a clamp-type pressure cell with silicone oil as a pressure medium. A Sn manometer was included in the teflon cup together with the specific-heat sample and its inductively determined superconducting transition temperature precisely measured pressure at low temperatures (33). An ac calorimetric technique (34) was used to measure heat capacity under pressure. A field-calibrated heater made of constantan wire was attached to one face of a plate-like sample, while a Cr-AuFe(0.07 %) thermocouple was glued to the other face. The heat capacity was obtained by measuring the ac voltage across the thermocouple wires, which is a direct measure of the ac temperature oscillation T$_{ac}$ incurred by ac heating. Heat capacity in arbitrary units was normalized by adiabatic measurements (11) to convert to absolute values.

The authors thank C. D. Batista, D. Pines, and N. J. Curro for discussions. Work at Los Alamos National Laboratory was performed under the auspices of the US Department of Energy, Office of Science, and supported by the Los Alamos Laboratory Directed Research and Development program.

**Figure Legends**

Figure 1. (a) Temperature-pressure phase diagram of $CeRhIn_5$ constructed from representative zero-field specific heat measurements plotted in (b). $T_N$ is the temperature at which long-range, local-moment antiferromagnetic (AFM) order develops and $T_c$ is the superconducting (SC) transition temperature. P1 is the pressure where $T_N$ is equal to $T_c$ and above which there is no evidence for AFM in zero-field measurements. P2 is the critical pressure where the projected $T_N(P)$ transition (dashed line) reaches zero. Dotted lines are guides to eyes. (b) Specific heat divided by temperature as a function of temperature for $CeRhIn_5$ at several pressures below and above P1. Arrows mark $T_c$ for each pressure. For P<P1, the higher temperature peak in C/T signals AFM order. The sharp, well-

defined SC and AFM transitions indicate that both are intrinsic bulk properties of $CeRhIn_5$ and rule out any significant pressure inhomogeneity as their origin. The specific heat data presented in this work are obtained by ac calorimetry (9).

Figure 2. (a) Specific heat divided by temperature for $CeRhIn_5$ at 1.4 GPa with 0 (circles) and 1.1-T (squares) magnetic fields. At 1.1 T, SC is suppressed below 300 mK. The solid line through these 1.1-T data and extending to T=0 is given by $\gamma_N + \eta T^2$, where the $T^2$ contribution is due to antiferromagnetic magnons in the AFM state and $\gamma_N$ is a measure of the electronic density of states in the absence of SC. These relatively large values of $\gamma_N$ are typical of strongly correlated U-based antiferromagnets (12). SC state values of C/T at T=0 K are labelled $\gamma(0)$. In zero applied field, C/T below 400 mK is extrapolated such that entropy is conserved below and above $T_c$ and the specific heat follows a $T^2$ dependence in the SC state (11). When AFM and SC coexist, the extrapolated $\gamma(0)$ in the SC state does not go to zero. It is not possible to determine the topology of gap nodes from these experiments, but the gap structure is expected to be complex because of coexisting incommensurate antiferromagnetism, which, by symmetry arguments, should produce a mixture of spin singlet and spin triplet superconducting electron pairs (13). The inset shows the superconducting transition in zero field after subtracting a magnetic background, i.e., $\Delta C=C(0T)-C(1.1T)$. (b) Specific heat jump $\Delta C$ at $T_c$ normalized by the product $\gamma_N T_c$ as a function of pressure. Inset: The electronic contribution to specific heat is plotted against pressure. Solid squares and circles denote $\gamma_N$ and $\gamma(0)$, respectively, from our specific heat measurements. Open symbols are from Ref. 11 and squares with inscribed crosses are estimated values of $\gamma_N$ from dHvA results (8). For P>P1, solid triangles denote the value of C/T just above $T_c$. By entropy conservation, these symbols define a lower bound on the T=0 K values of $\gamma_N$.

Figure 3. Field-temperature phase diagram determined from specific heat measurements with fields applied perpendicular to the c-axis of CeRhIn$_5$ at (a) P=1.51 GPa, (b) 1.62 GPa, and (c) 1.71 GPa. $B_N$ is a magnetic transition to an incommensurate magnetic structure (ICM) with propagation vector (1/2,1/2,0.297) (circles); $B_M$ is a spin reorientation transition to a commensurate (1/2,1/2,1/4) magnetic structure (CM) (squares) (ref. 18); $B^*$ is a magnetic transition to an incommensurate phase (ICM2) that is observed in a limited field and temperature range; and $B_{c2}$ denotes the upper critical field required to suppress superconductivity. Solid lines are guides to eyes. In zero field, unconventional superconductivity, which involves partially itinerant 4f electrons, coexists with incommensurate antiferromagnetic order. Neutron-diffraction experiments find that the magnetic structure is unchanged and ordered magnetic moment is reduced only slightly at these pressures compared to its value at atmospheric pressure (14).

Figure 4. (a) Upper critical field normalized by the zero-field $T_c$ ($B_{c2}/T_{c0}$) versus normalized temperature ($T/T_{c0}$), where the superconducting transition temperature $T_{c0}$ is 0.9, 1.12, 1.41, 2.27, and 2.18 K for 1.4, 1.51, 1.62, 2.1, and 2.45 GPa, respectively. At 2.1 GPa, the half-filled diamonds denote a phase boundary for field-induced magnetism that develops inside the SC phase and extends above $B_{c2}$ (ref. 9). Solid lines are guides to eyes. For P>P1, the T=0 value of the orbital field, estimated from the initial slope at $T_{c0}$ ($B^{orb}_{c2}(T=0)=0.73T_c\partial B_{c2}/\partial T$), is much greater than the measured $B_{c2}$ at low temperatures, indicating strong Pauli paramagnetic limiting. Values of the upper critical field slope are plotted in (b). The absence of Pauli limiting below P1 could be due to a spin-triplet component in superconducting electron pairs, which disappears when AFM disappears above P1. (b) Pressure dependence of the Fermi velocity $v_F$, plotted on the left ordinate, and of the derivative of the upper critical field at $T_c$, $\partial B_{c2}/\partial T$ (=$B_{c2}'$), plotted on the right ordinate. $v_F$ is estimated from the initial slope of $B_{c2}$ (circles) for clean-limit singlet superconductors appropriate to CeRhIn$_5$: $B_{c2}' = (2.54 \times 10^8 \text{ Tm}^2\text{K}^{-2}\text{sec}^{-2}) T_c / v_F^2$ (ref. 20).

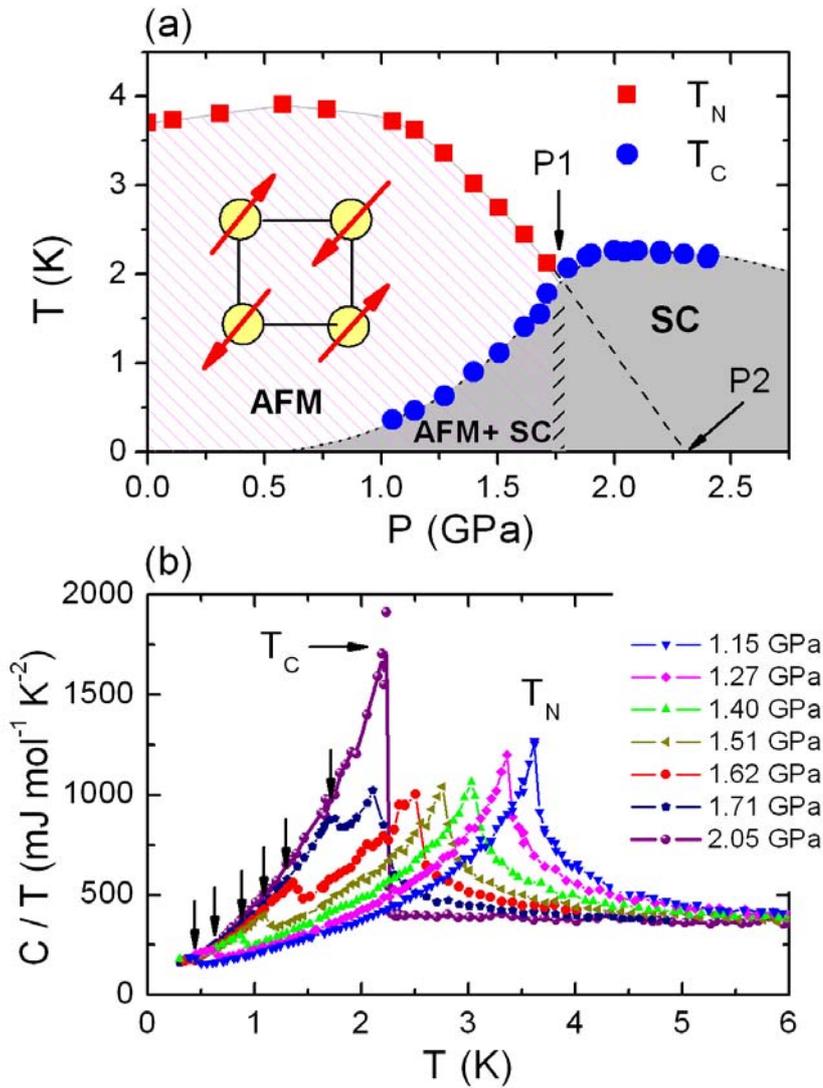

Figure 1

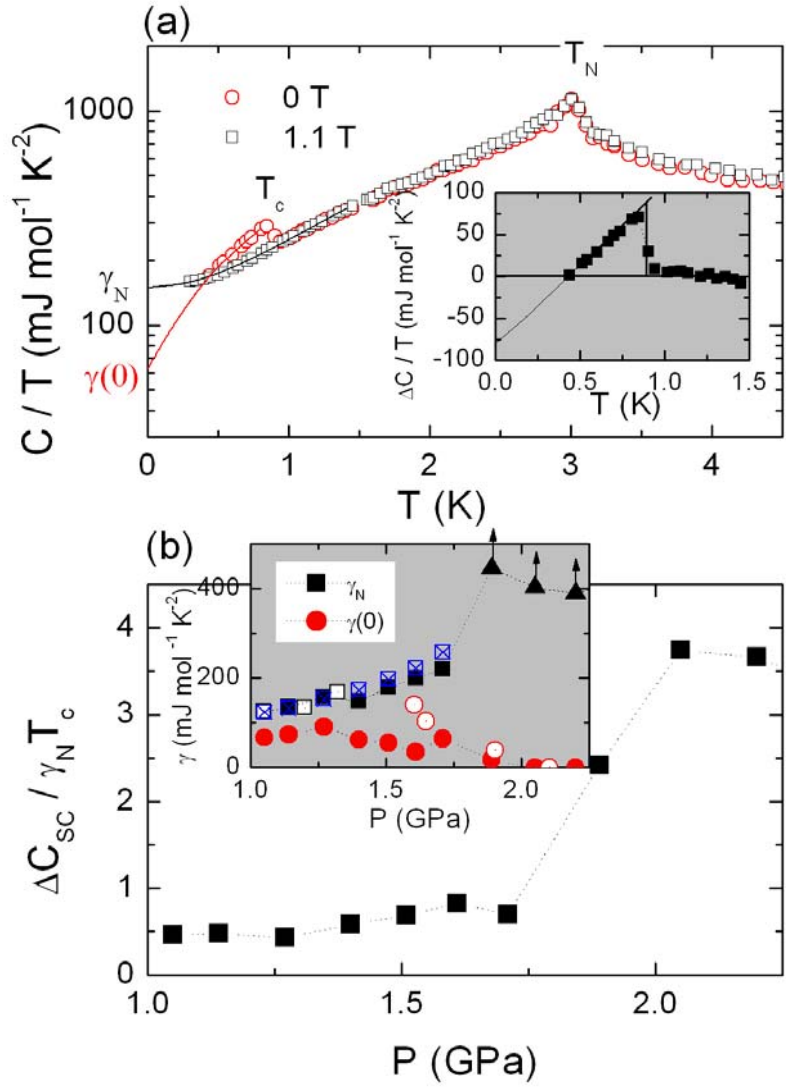

Figure 2

- 13 -

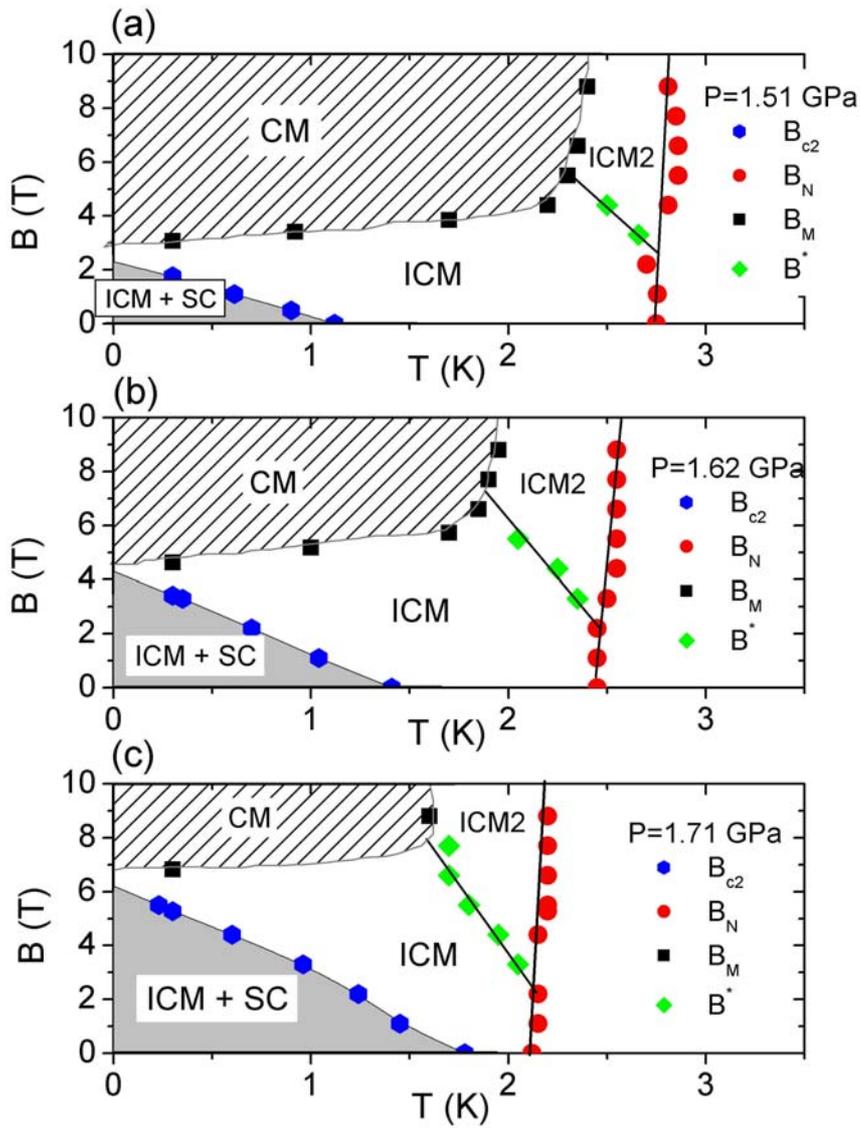

Figure 3

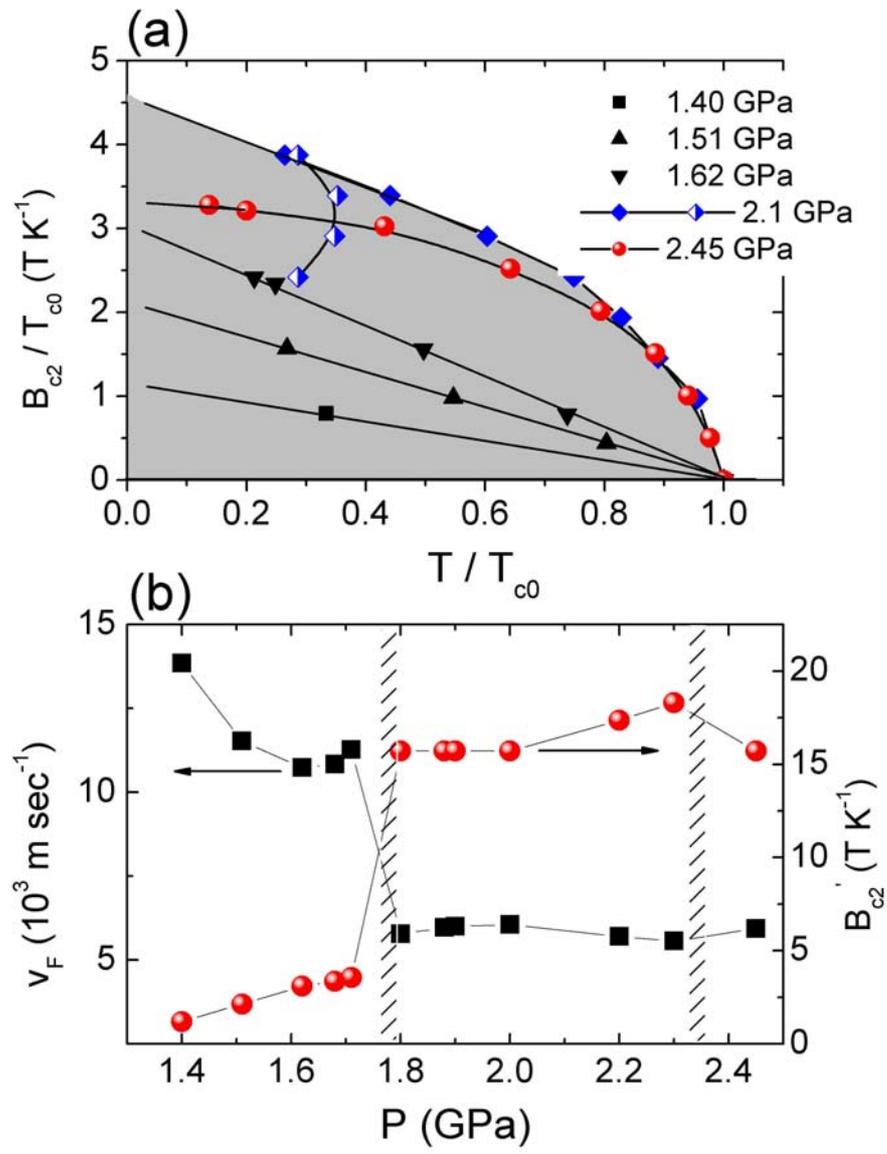

Figure 4